\pdfoutput=1

\documentclass[11pt,twoside,a4paper,cmspaper,final,collab]{cms-tdr}
\def\svnVersion{177384}\def\cmsCernNo{2012-206}\def\cmsCernDate{\today}\def\cmsMessage{Submitted to Physics Letters B}

\begin{document}\cmsNoteHeader{TOP-11-028}

\hyphenation{had-ron-i-za-tion}
\hyphenation{cal-or-i-me-ter}
\hyphenation{de-vices}

\RCS$Revision: 160530 $
\RCS$HeadURL: svn+ssh://svn.cern.ch/reps/tdr2/papers/TOP-11-028/trunk/TOP-11-028.tex $
\RCS$Id: TOP-11-028.tex 160530 2012-12-07 16:44:09Z yuanchao $
\newlength\cmsFigWidth
\ifthenelse{\boolean{cms@external}}{\setlength\cmsFigWidth{0.85\columnwidth}}{\setlength\cmsFigWidth{0.4\textwidth}}
\ifthenelse{\boolean{cms@external}}{\providecommand{\cmsLeft}{top}}{\providecommand{\cmsLeft}{left}}
\ifthenelse{\boolean{cms@external}}{\providecommand{\cmsRight}{bottom}}{\providecommand{\cmsRight}{right}}
\providecommand{\met}{\ETslash}
\providecommand{\st}{\ensuremath{\mathrm{S}_\mathrm{T}}\xspace}
\providecommand{\rj}{\ensuremath{\mathrm{j}}}
\cmsNoteHeader{TOP-11-028} 
\title{Search for flavor changing neutral currents in top quark decays in $\Pp\Pp$ collisions at 7\TeV}

\date{\today}

\abstract{
The results of a search for  flavor changing neutral currents in top quark  decays $\cPqt\to  \cPZ\cPq$
in  events with a topology compatible with the decay chain
$\ttbar \to \PW\cPqb+\cPZ\cPq \to  \ell\nu \cPqb +\ell\ell \cPq$ are presented. The search is
performed with  a data
sample corresponding to an integrated luminosity of 5.0\fbinv of proton-proton collisions
at a center-of-mass energy of 7\TeV, collected with  the CMS detector at the LHC.
The observed number of events agrees with the standard model prediction and no
evidence for flavor changing neutral currents in top quark decays is found.
A $\cPqt \to \cPZ\cPq$ branching fraction
greater than 0.21\%  is excluded at the 95\% confidence level.
}

\hypersetup{%
pdfauthor={CMS Collaboration},%
pdftitle={Search for flavor changing neutral currents in top quark decays in pp collisions at 7 TeV},%
pdfsubject={top},%
pdfkeywords={CMS, physics, top, searches}}

\maketitle 

\section{Introduction}
\label{sec:introduction}
\par
The top quark decays with a branching fraction of nearly 100\% to a bottom quark and  a $\PW$ boson,
$\cPqt \to  \PW\cPqb$. However, some extensions of the  standard model\,(SM)
 predict that the top quark can also decay through a neutral
$\cPZ$ boson, $\cPqt \to  \cPZ\cPq$, where $\cPq$ is a $\cPqu$ or $\cPqc$ quark.
This decay is suppressed in the SM by the GIM mechanism~\cite{ref:GIM} and  occurs
at the level of quantum loop corrections only. The branching fraction $\mathcal{B}(\cPqt \to  \cPZ\cPq)$
is predicted to be $\mathcal{O}(10^{-14})$~\cite{ref:Glover:2004cy}, far below the experimental
reach of the Large Hadron Collider\,(LHC).
Detection of this signal would therefore be an indication
of a large enhancement in the branching fraction and clear evidence    for  violations of the SM prediction.
There are several models, for example R-parity-violating supersymmetric models~\cite{AguilarSaavedra:2000db}
and topcolor-assisted technicolor models~\cite{Lu:2003yr}, that predict enhancements of the $\cPqt \to \cPZ\cPq$ decay
where  $\mathcal{B}(\cPqt \to  \cPZ\cPq)$ could be as large as $\mathcal{O}(10^{-4})$.

Previous searches for the flavor changing neutral currents in top quark decays performed at the Tevatron by
CDF and D0  determined
a $\mathcal{B}(\cPqt \to  \cPZ\cPq)$ upper limit of 3.7\%~\cite{ref:2008aaa}  and 3.2\%~\cite{ref:Abazov:2011qf}
at the 95\% confidence level\,(CL), respectively.
At a center-of-mass energy of 7\TeV, the $\ttbar$  production cross section
at the LHC at the next-to-leading order is 157.5\unit{pb} for an assumed top quark mass of 172.5\GeV,
which   is
twenty times larger than that  at the Tevatron at a center-of-mass energy of 2\TeV.
This enables event samples with leptonically decaying vector bosons to be used more effectively.
These samples have well determined backgrounds.  A recent search
in the three-lepton channels performed at ATLAS with an integrated luminosity of 2.1\fbinv
reported  a
$\mathcal{B}(\cPqt \to \cPZ\cPq) $    upper limit
of 0.73\%~\cite{ref:ATLAS-2012}.

We  expect ${\cal B}(\cPqt \to \cPZ\cPq)$ to be  small and
look for ${\ttbar \to   \cPZ\cPq + \PW\cPqb \to  \ell\ell \cPq +  \ell} \nu\cPqb$ final state events,  which produce
three-lepton ($\Pe\Pe\Pe, \Pe\Pe\mu, \mu\mu\Pe, \mu\mu\mu$) final states.
This choice   results in a measurement with reduced background  and fewer signal events.
The analysis uses a data sample corresponding to an integrated
luminosity of 5.0\fbinv
of proton-proton collisions at $\sqrt{s} = 7$\TeV,
recorded by the Compact Muon Solenoid\,(CMS)  experiment during  2011.

\section{The CMS Detector}
\label{sec:cms}
The central feature of the CMS apparatus is a superconducting solenoid, 13\unit{m} in length and 6\unit{m} in
diameter, which provides an axial magnetic field of 3.8\unit{T}. Within the field volume there are several
particle detection systems. Charged particle trajectories are
measured by silicon pixel and silicon strip trackers, covering $0\le \phi \le 2\pi$ in azimuth and
$\abs{\eta}  < 2.5$  in pseudorapidity,  where $\eta$ is
defined as $-\log[\tan \theta/2]$ and    $\theta$ is the polar angle of the trajectory of the
particle with respect to the
counterclockwise proton beam direction. A crystal electromagnetic calorimeter and a brass/scintillator
hadron calorimeter surround
the tracking volume, providing energy measurements of photons,
electrons and hadron jets. Muons are
identified and measured in gas-ionization
detectors embedded in the steel return yoke outside the solenoid. The detector is nearly hermetic,
allowing energy balance measurements in
the plane transverse to the beam direction. A two-tier trigger system selects the most interesting
proton-proton collision events for use in physics
analysis. A more detailed description of the CMS detector can be found in Ref.~\cite{ref:cms}.

\section{Basic Selection}
\label{sec:preselection}

Events with two opposite-sign, isolated leptons ($\Pe$ or $\mu$)  consistent with a $\cPZ$-boson decay and an extra charged
lepton are selected, $\Pep\Pem\Pe^\pm, \Pep\Pem\mu^\pm, \Pgmp\Pgmm\Pe^\pm, \Pgmp\Pgmm\Pgm^\pm$. All three leptons
must be isolated and have  transverse momentum {$\pt>20$}\GeV, and the electrons
(muons) must have $\abs{\eta} < 2.5~(\abs{\eta} < 2.4)$.
Events are required to pass at least one of the $\Pe\Pe$ or $\mu\mu$ high-$\pt$ double-lepton triggers. Their
efficiencies for events containing
two leptons satisfying the analysis selection are  measured to be  99\%, 98\%, 91\% and 93\%
for the  $\Pe\Pe\Pe$, $\Pe\Pe\mu$, $\mu\mu\Pe$ and  $\mu\mu\mu$  channels, respectively.

Muon candidates are reconstructed with a global fit of trajectories using hits in the tracker and the muon system.
The muon candidate must have associated hits in the silicon strip and pixel detectors, have
segments in the muon chambers, and have a high-quality global fit to the track trajectory. The efficiency for these
muon selection criteria is at least $99$\%~\cite{Khachatryan:2010xn}.

Electron reconstruction starts from clusters of energy deposits in the electromagnetic calorimeter, which are
matched to hits in the silicon strip and the pixel detectors.
Electrons are identified using variables which include the ratio between the energy deposited in the hadron and
the electromagnetic calorimeters, the shower width in
$\eta$, and the distance between the calorimeter shower and the particle trajectory in the tracker, measured
in both $\eta$ and $\phi$.     The selection criteria used are optimized~\cite{Khachatryan:2010xn} to maintain
an efficiency of
approximately 95\% for the electrons from $\PW$ or $\cPZ$ decays.

The invariant mass of at least one  $\Pep\Pem$ or $\Pgmp\Pgmm$  pair is required to be between
60\GeV and 120\GeV. If two dilepton pairs lie in this mass window, the one closest to the $\cPZ$ mass is taken.
Due to the high instantaneous luminosity of the LHC, there are multiple interactions per bunch  crossing (pileup). Therefore,
events are required to have at least one  good  primary vertex,
which is chosen as the vertex with the highest $\Sigma {\pt}^2$
of its associated tracks.
All leptons, which are used to select or reject events, must come from the same primary vertex.
The $\Pgmp\Pgmm$  pair  opening angle is required to differ from $\pi$ radians
by more than 0.05 radians to reject cosmic rays.

Electrons and muons from $\cPZ$ and $\PW$ decays are expected to be isolated from other particles. A cone of size
$\DR \equiv \sqrt{(\Delta\eta)^2 + (\Delta \phi)^2}=0.3$  is constructed
around the lepton momentum direction.  The lepton relative isolation is quantified by summing the transverse energy
(as measured in the calorimeters) and the
 transverse momentum (as measured in the silicon tracker) of all objects within this cone,
excluding the lepton, and then dividing
by the lepton transverse momentum~\cite{Chatrchyan:2012xi}. The resulting quantity, corrected
for additional underlying event activity due to
pileup events, is required to be less than 0.125 (0.1) for
$\cPZ \to  \ell^+\ell^-$ ($\PW \to \ell\nu$).  This requirement
rejects misidentified leptons and  background arising  from hadronic jets.

The third lepton in the event should be the result of a leptonic decay of a $\PW$ boson.
In order to increase the electron purity, more stringent reconstruction requirements are used  for
$\PW\to \Pe\nu$ candidates.  In this case the selection criteria are optimized~\cite{Khachatryan:2010xn} to reject
the background from  jets while maintaining an efficiency of  80\% for the electrons from $\PW$ or $\cPZ$ decays.
The muon purity for the $\cPZ$ selection described above is high and the same reconstruction requirements are
 used to identify ${\PW\to}\mu\nu$ candidates.
Events with a fourth lepton satisfying the $\PW\to \ell\nu$ criteria are rejected.

The jets and the missing transverse  energy vector ($-\Sigma \vec{p}_{\mathrm{T}}$)  and its magnitude ($\met$)
are reconstructed using a particle-flow technique~\cite{ref:pf}.
An  anti-\kt clustering algorithm~\cite{ref:kt} with a distance parameter of 0.5 is used for jet reconstruction.
The energy calibration~\cite{ref:jetscale}
is performed separately for each particle type in the jet,
and the resulting jet energies require only a small correction accounting
for thresholds and residual inefficiencies.
In addition, a correction for pileup is included and
jets are required to satisfy identification criteria that eliminate
jets originating from noisy channels in the calorimeters~\cite{ref:met,ref:noise}.
Jets are required to have  $\pt > 30$\GeV,  $\abs{\eta} < 2.4$, and to
be  separated by $\DR > 0.4$ from leptons passing the analysis selection.
Neutrinos from $\PW$-boson decays escape detection and produce a significant momentum imbalance in the detector.
We require  the missing transverse  energy  to be larger than 30\GeV.

The samples of Drell--Yan events
with invariant mass of lepton pairs $ m_{\ell\ell}$ larger than  50\GeV,
SM $\ttbar$, $\cPZ \ttbar$,
$\PW \ttbar$ and $\PW\cPZ$ are generated using $\MADGRAPH$~\cite{ref:madg}.
The samples of  $\PW\PW$  and  $\cPZ\cPZ$ diboson events are simulated using
\PYTHIA~\cite{ref:pythia}, while single-top-quark events are generated using
\POWHEG~\cite{ref:Nason:2004rx,ref:Frixione:2007vw,ref:Alioli:2010xd}.
The signal sample  $\Pp\Pp \to \ttbar \to   \cPZ\cPq + \PW\cPqb \to$
$\ell^+\ell^- \cPq + \ell^\pm\nu\cPqb\ ( \ell = \Pe , \mu, \tau)$
is generated with \MADGRAPH and the
top quarks decay and hadronize through \PYTHIA.
Due to the loss of top quark spin information for FCNC in \PYTHIA, events are
reweighted according to the SM prediction of the helicity distribution.
This study is not sensitive to the choice of anomalous coupling settings, which
are taken into account in systematic uncertainties.
The set of parton distribution functions used is CTEQ6L~\cite{ref:CTEQ6L}.  The CMS detector response is
simulated using a $\GEANTfour$-based~\cite{ref:geant} model, and the events are
 reconstructed and analyzed using the same software used to process collision data.
The  simulated events are weighted so that the trigger efficiencies, reconstruction efficiencies  and the distribution
of reconstructed vertices observed in data are reproduced.

The observed and expected yields based on MC after the basic event selection described above are listed in Table~\ref{tab:preselection}. The
initial data  sample of 1.3\,(1.6) million $\cPZ$
to $\Pe\Pe\,( \mu\mu )$
 events is reduced to less than 100 events per three-lepton channel.
 All  entries in Table~\ref{tab:preselection}
 also include the $\tau$ decay mode contributions.
Single-top-quark production is dominated
by the $\PW\cPqt$ channel. The total yields are dominated by diboson production  and a reasonable agreement is observed
between data and simulation.
The details of the background estimations are discussed in Section~\ref{sec:back}.

\begin{table*}[htb]
\topcaption{
\label{tab:preselection}
Event yields and background predictions based on simulated events for all three-lepton channels  after
the  basic event selection, which includes
the trigger,  $\cPZ$ boson, third lepton, fourth-lepton veto and    missing transverse  energy  requirements
for an integrated luminosity of 5.0\fbinv.
The uncertainties include the statistical and systematic components separately (in that order).
}
{\small
\begin{center}
\begin{tabular}{|l| c c c c |}
\hline
\hline
Channel      & $\mu\mu\Pe$ & $ \mu\mu\mu$ & $\Pe\Pe\Pe$   & $\Pe\Pe\mu$   \\
\hline
\hline
Drell--Yan   &  2.0 $\pm$ 0.9$\pm$ 0.3  &  0.9 $\pm$ 0.6 $\pm$ 0.1   &  2.8 $\pm$ 1.1 $\pm$ 0.4  &  0.9 $\pm$ 0.6 $\pm$ 0.1       \\
${\PW\cPZ}$      &  46.1 $\pm$ 0.3 $\pm$ 6.1  &  60.3 $\pm$ 0.4 $\pm$ 8.0 &  40.9 $\pm$ 0.3 $\pm$ 5.4  &  48.6 $\pm$ 0.4 $\pm$ 6.4   \\
${\cPZ\cPZ}$      &  17.7 $\pm$ 0.2 $\pm$ 2.3  &  21.7 $\pm$ 0.2 $\pm$ 2.9  &  15.1 $\pm$ 0.2 $\pm$ 2.0 &  18.2 $\pm$ 0.1 $\pm$ 2.4   \\
$\cPZ\ttbar$ &  2.2  $\pm$ 0.1 $\pm$ 1.4  &  2.4 $\pm$ 0.1 $\pm$ 1.5  &  2.0 $\pm$ 0.1 $\pm$ 1.2  &  2.3 $\pm$ 0.1 $\pm$ 1.4   \\
$\PW\ttbar$ &  0.31 $\pm$ 0.02 $\pm$ 0.21 & 0.26 $\pm$ 0.02 $\pm$ 0.18 & 0.21 $\pm$ 0.02 $\pm$ 0.14 & 0.29 $\pm$ 0.02 $\pm$ 0.20 \\
${\PW\PW}$    & $\le$ 0.001  & $\le$ 0.001  &  0.18 $\pm$ 0.06 $\pm$ 0.01 & $\le$ 0.001             \\
$\ttbar$ & $\le$ 0.001  &  0.5 $\pm$ 0.2 $\pm$ 0.1  &  0.9 $\pm$ 0.5 $\pm$ 0.1  &  0.9 $\pm$ 0.4 $\pm$ 0.1   \\
Single top   & $\le$ 0.001  &  0.14 $\pm$ 0.09 $\pm$ 0.02 &  $\le$ 0.001  &  $\le$ 0.05   \\
\hline
\hline
Total        &  69 $\pm$ 1 $\pm$ 7 &  86 $\pm$ 1$\pm$ 9 &  62 $\pm$ 1 $\pm$ 6     &  72 $\pm$ 1 $\pm$ 7   \\
\hline
Observed     &  73   &  87 &  85 &  61\\
\hline
\hline
\end{tabular}
\end{center}
}
\end{table*}

\begin{figure*}[htb!]
\begin{center}
\includegraphics[width=0.45\textwidth]{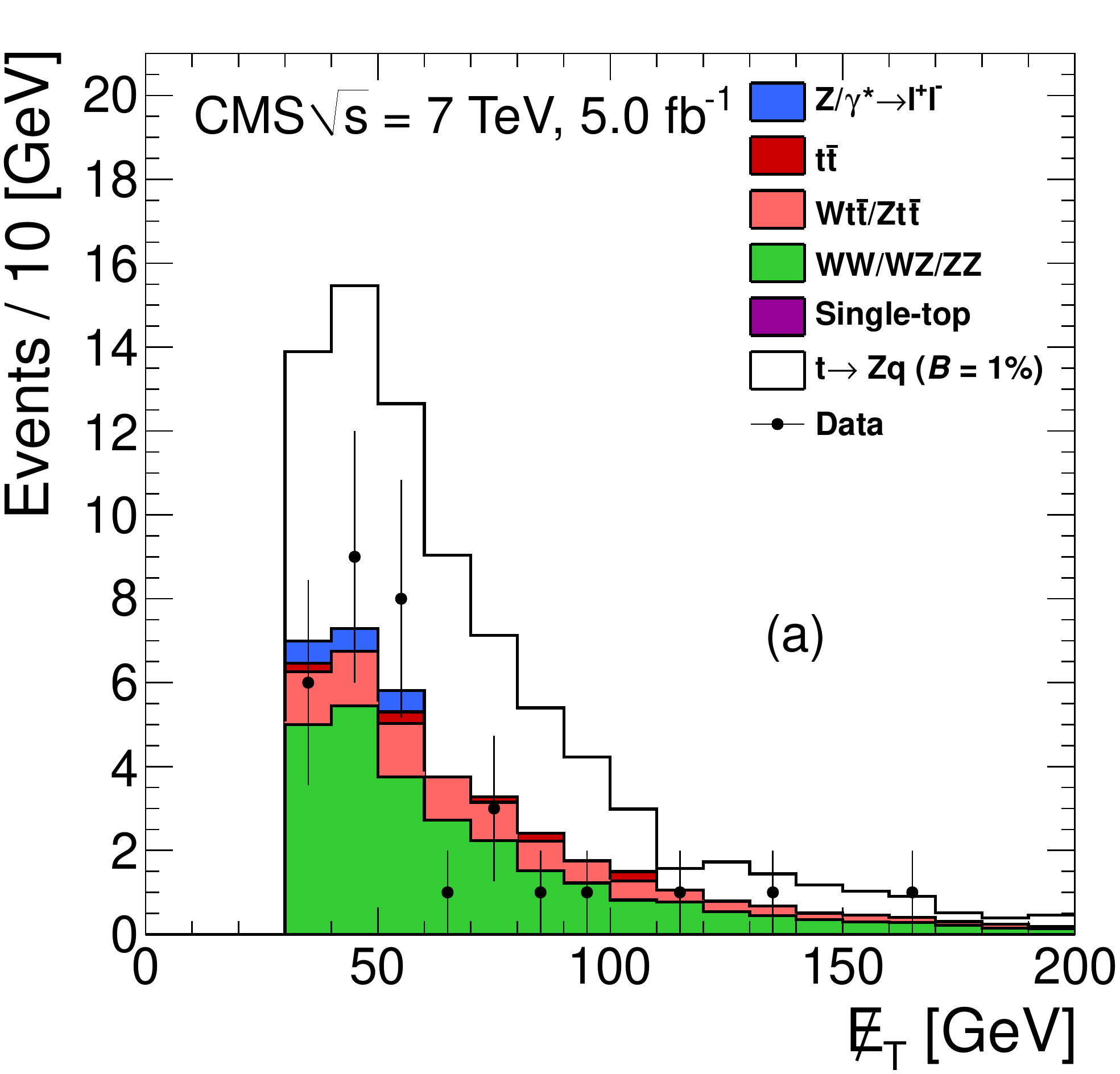}\\
\includegraphics[width=0.45\textwidth]{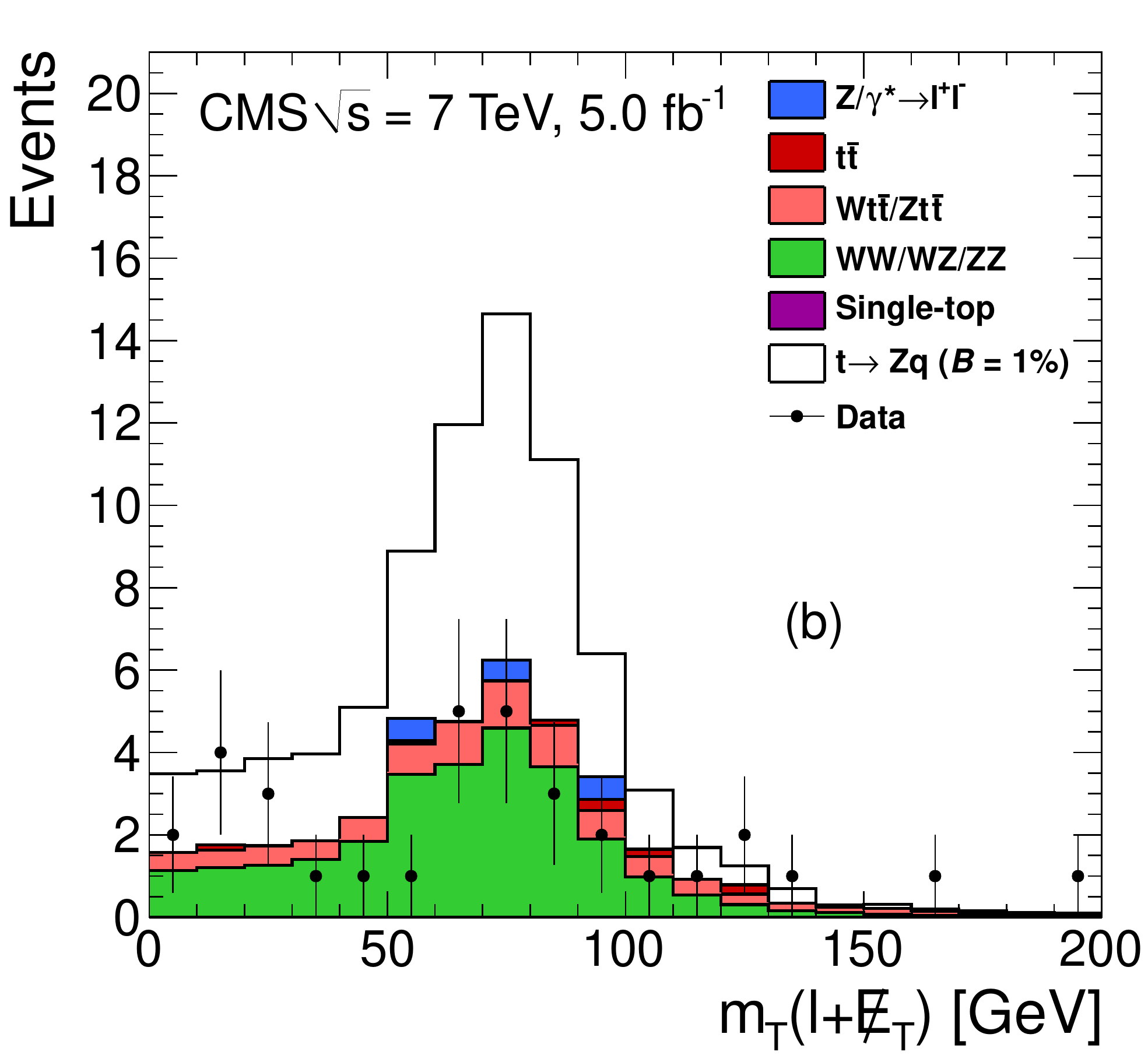}
\includegraphics[width=0.45\textwidth]{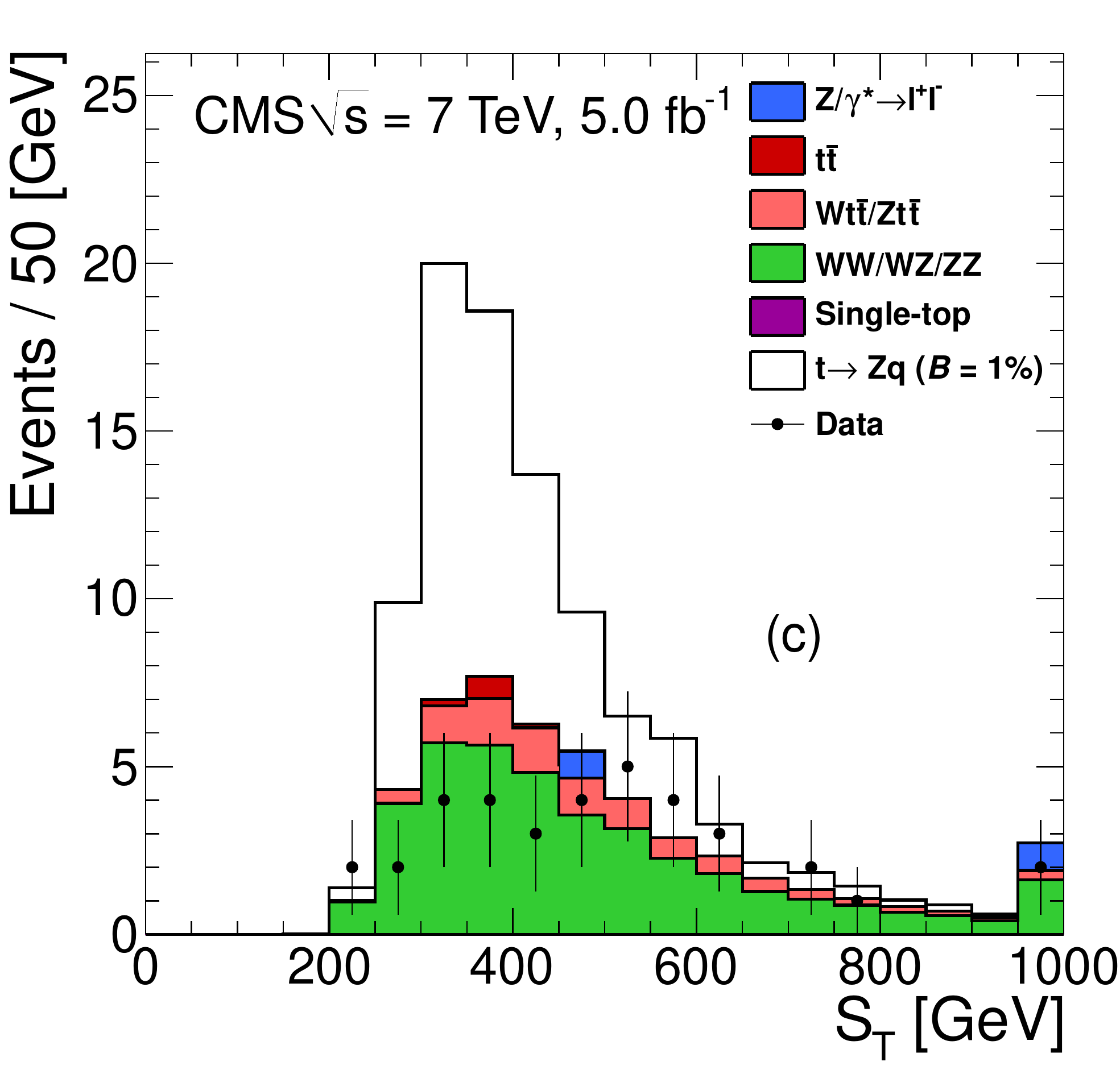}
\caption{\label{fig:jet} Comparison between data and simulated events
for an integrated luminosity of 5.0\fbinv,
after
the basic event selection described in Section~\ref{sec:preselection} and requiring at least
two jets,
for:
(a) the missing transverse energy distribution, (b) the reconstructed $\ell \nu$ transverse mass of the $\PW$
boson candidate, and (c) the scalar sum of the transverse energy for the jets, charged leptons,
and neutrino, \st.
The data are represented by the points with error bars and the open histogram shows  the expected signal
assuming $\mathcal{B}(\cPqt\to \cPZ\cPq)$ is equal to 1\%.
Stacked solid histograms represent the dominant backgrounds. The statistical
uncertainties of these backgrounds are around a few percent level and are
not drawn.
}
\end{center}
\end{figure*}

Figure~\ref{fig:jet} shows  the distributions for data and simulated events of
the missing transverse energy, transverse mass of the $\PW$ boson candidate ($m_\mathrm{T}$),
and the scalar sum of the transverse energy
\st, after the trigger,  $\cPZ$ boson, third lepton, fourth-lepton veto, missing transverse energy,
and the additional requirement of two or more jets. The \st variable
is defined as $ \Sigma { p_{\mathrm{T}\ell}} + \Sigma { p_{\mathrm{T}\rj}} + \met$, where only the three leptons and
two jets from the $\ttbar$ candidate are considered. The $m_\mathrm{T}$ is calculated using the transverse
momentum and azimuthal direction
of the third  lepton  and the magnitude and direction of the missing transverse energy, as
$\sqrt{2p_{\mathrm{T}\ell}\met(1-\cos(\Delta\phi))}$.

\section{Signal Reconstruction}
\label{sec:signal}
For the  $\cPqt \to  \cPZ\cPq \to \ell^+\ell^- \rj$ signal,  a full reconstruction of the top quark mass $m_{\cPZ \rj}$
  is possible and  straightforward,  but the possibility of a combinatorial background arises since
there is no unambiguous way to pair multiple  light-quark jets with the $\cPZ$ boson. Therefore all possible
 combinations are examined.

The invariant mass of the W and b jet system ($m_{\PW\cPqb}$) can be reconstructed by assuming that the
transverse components  of the neutrino momentum are given by the missing transverse energy vector information,
while the  longitudinal component is  calculated as
\ifthenelse{\boolean{cms@external}}{
\begin{multline*}
 p_{z\nu}= \frac{p_{z\ell}(p_{x\ell} p_{x\nu}+p_{y\ell} p_{y\nu}+{ m_\PW}^2/2)}{{E_\ell^2-p_{z \ell}^2}}\\
\pm\frac{E_\ell \sqrt{(p_{x\ell} p_{x\nu}+p_{y\ell} p_{y\nu}+{m_\PW}^2/2)^2-E_{\mathrm{T}\nu}^2(E_\ell^2-p_{z\ell}^2)}} {E_\ell^2-p_{z \ell}^2},
\end{multline*}
}
{
\begin{equation*}
 p_{z\nu}= \frac{p_{z\ell}(p_{x\ell} p_{x\nu}+p_{y\ell} p_{y\nu}+{m_\PW}^2/2)\pm
E_\ell \sqrt{(p_{x\ell} p_{x\nu}+p_{y\ell} p_{y\nu}+{m_\PW}^2/2)^2-E_{\mathrm{T}\nu}^2(E_\ell^2-p_{z\ell}^2)}} {E_\ell^2-p_{z \ell}^2},
\end{equation*}
}
where ${E_{\ell}}$, ${p_{x \ell}}$, ${p_{y\ell}}$,  and ${p_{z\ell}}$ are the energy and momentum
components for the lepton, while the neutrino
 ${E_{\mathrm{T}\nu}}$, ${p_{x \nu}}$  and ${p_{y\nu}}$ are estimated from the reconstructed  missing transverse
energy  magnitude and direction, and imposing the constraint that the invariant mass of the lepton and
the neutrino is equal to the $\PW$-boson mass ($m_\PW$).  If the discriminant is found to be negative, it is set equal to
zero. In events in which there are two possible solutions for $p_{z\nu}$, the solution with the smaller magnitude of $p_{z\nu}$,
 is taken; studies
with simulated signal events show that this solution is the correct one more
than 60\% of the time.

Next, we add the requirements on jets, $m_{\cPZ \rj}$, and $m_{\PW\cPqb}$ to the basic selection described
in Section~\ref{sec:preselection}, and search for $\ttbar\to \PW\cPqb + \cPZ\cPq$ in two ways.
One selection requires a minimum  value of \st and  loose requirements on $m_{\cPZ \rj}$ and $m_{\PW\cPqb}$.
The second selection is stricter,
with tight requirements on the $m_{\cPZ \rj}$ and $m_{\PW\cPqb}$ quantities and the requirement
that one of the jets should
be consistent with the hadronization of a $\cPqb$ quark, namely a ``\cPqb\ jet''.
In this Letter, we refer to these two selections as the ``\st'' and ``$\cPqb$-tag'' selections, respectively.
The first selection is the more sensitive and hence is taken as the reference analysis.
Table~\ref{tab:effy} shows the estimates of the overall signal efficiency determined from simulated events.

\begin{table}[htb]
\begin{center}
\topcaption{
\label{tab:effy}
Signal selection efficiency  for each three-lepton channels in percent.  The efficiency is calculated as the fraction of events
with leptonically (${\Pe,}\,\mu,\,\tau$) decaying
$\PW$ and $\cPZ$ bosons  passing the selection.
Only statistical uncertainties are shown.
}
\begin{tabular}{| l |  c | c|}
  \hline
  \hline
Channel & \st Selection $[\%]$   & $\cPqb$-tag Selection $[\%]$ \\
 \hline
  \hline
$\Pe\Pe\Pe$        & (12.4 $\pm$ 1.1)  & (3.8 $\pm$ 0.6)\\
$\Pe\Pe\mu$    & (13.8 $\pm$ 1.2)  & (5.0 $\pm$ 0.7)\\
$\mu\mu \Pe$   & (14.8 $\pm$ 1.2)  & (5.1 $\pm$ 0.7)\\
$\mu\mu\mu$      & (14.7 $\pm$ 1.2)  & (5.3 $\pm$ 0.7)\\
\hline
  \hline
\end{tabular}
\end{center}
\end{table}

\subsection{\texorpdfstring{\st}{S[T]} Selection}

In the \st selection,
at least two jets with  $\pt > 30$\GeV are required, which
are assumed to come from the primary vertex.
A constituent track candidate in a jet is removed from
the reconstruction if it does not point to the same vertex, but there is no
association requirement between jets and the $\cPZ$ candidate which is
chosen in the basic selection.

A candidate event is required to have \st above 250\GeV,
$m_{\cPZ \rj}$  and $m_{\PW\cPqb}$ are required to be between 100\GeV and 250\GeV.
The \st requirement reduces the boson-jet combinations.
The \st distribution of the best candidate is shown in
Figure~\ref{fig:jet}. All possible $\ttbar$ combinations are examined
and the reconstructed $\ttbar$ pair that has the largest separation in
azimuthal angle is selected.
Figure~\ref{fig:mass}\,(top) shows the comparison of the distributions of
$m_{\cPZ \rj}$ and $m_{\PW\cPqb}$ in data and simulation   after the basic event
selection described in Section~\ref{sec:preselection}
(Table~\ref{tab:preselection}), combined with
the  two  or more jets  and the  \st requirements.

\subsection{b-tag Selection}

To further reduce the background from diboson events,
a $\cPqb$-tag based selection is performed. In this selection, at least two
jets are required to be associated with the primary vertex associated with the  $\cPZ$ candidate and the event can contain only
one $\cPqb$ jet.

The $\cPqb$ jets  are identified by the track counting high-efficiency $\cPqb$-tagging algorithm described
in Ref.~\cite{ref:btag}, which relies on tracks with large impact parameter significance.
This tagging method has an identification efficiency of 65\% to 85\% for $\cPqb$ jets
with transverse momentum between 30\GeV to 100\GeV and a misidentification rate below 15\%.

The jet which gives the invariant mass of $m_{\cPZ \rj}$ closest to the top mass is selected and the reconstructed top quark mass $m_{\cPZ \rj}$ is required to be within 25\GeV  of the assumed
top quark mass,  $m_\cPqt = 172.5$\GeV, while $m_{\PW\cPqb}$  is required to be within 35\GeV  of $m_\cPqt$.

\begin{figure*}[htbp]
\begin{center}
\includegraphics[width=0.45\textwidth]{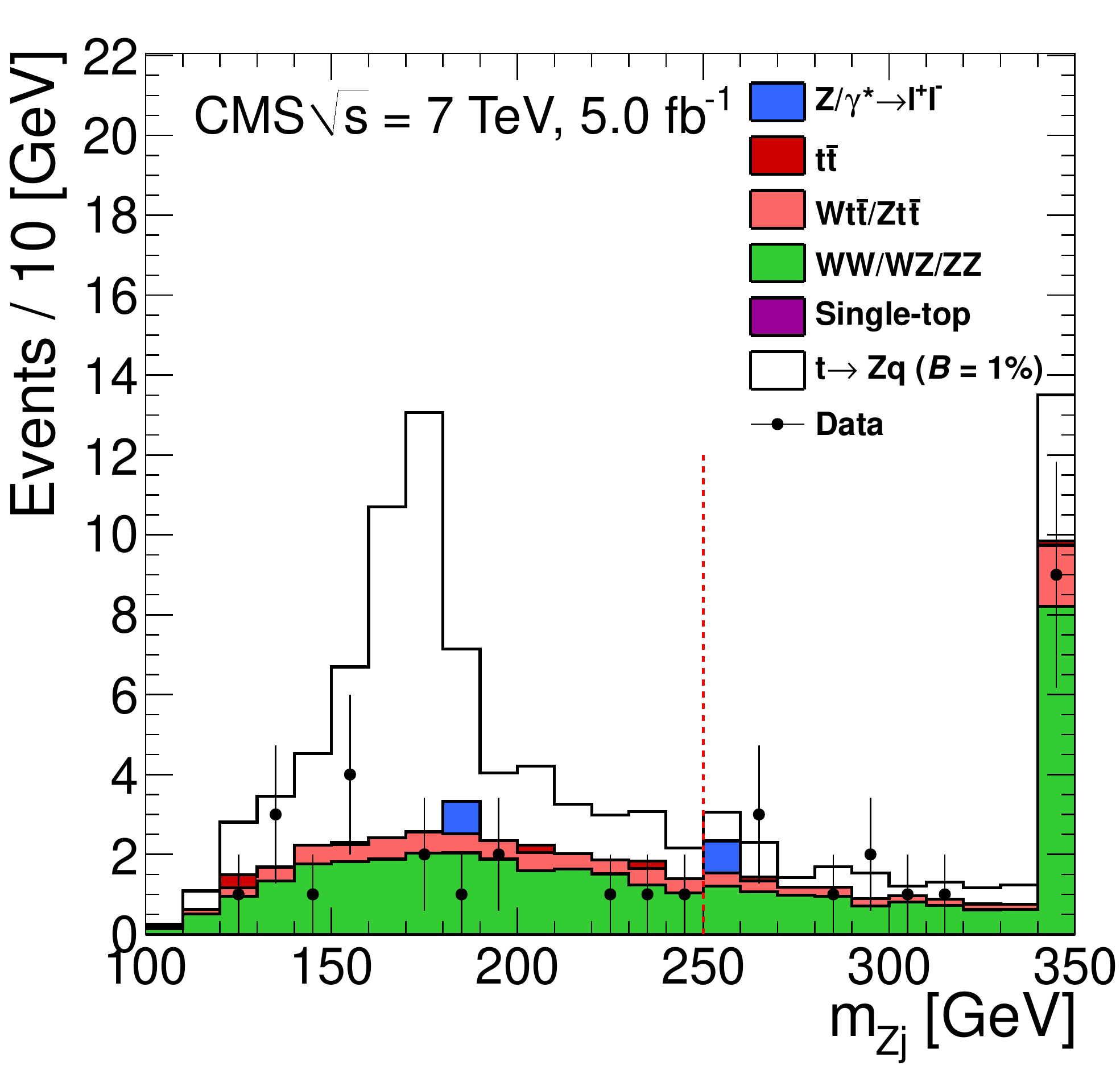}
\includegraphics[width=0.45\textwidth]{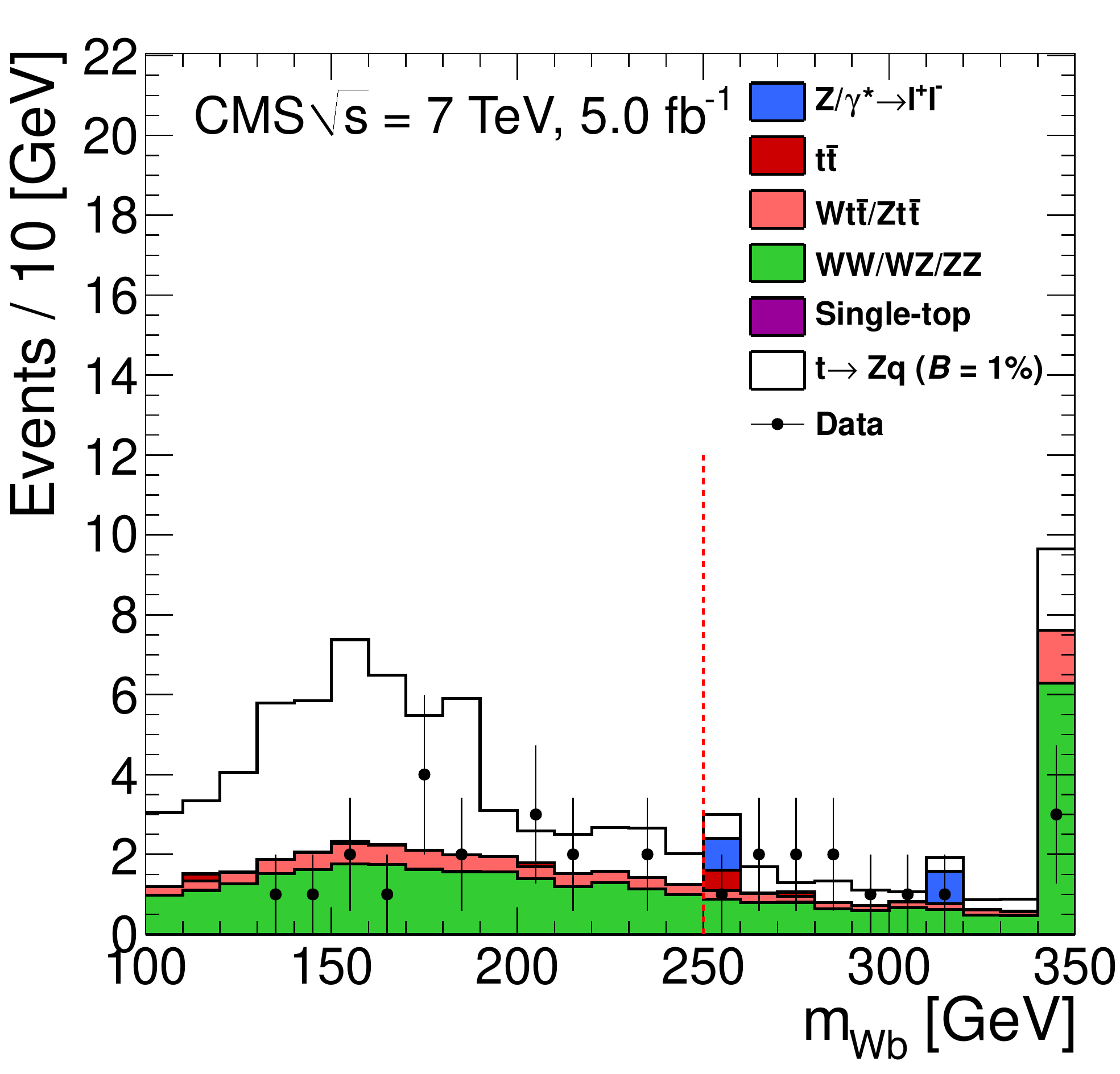}\\
\includegraphics[width=0.45\textwidth]{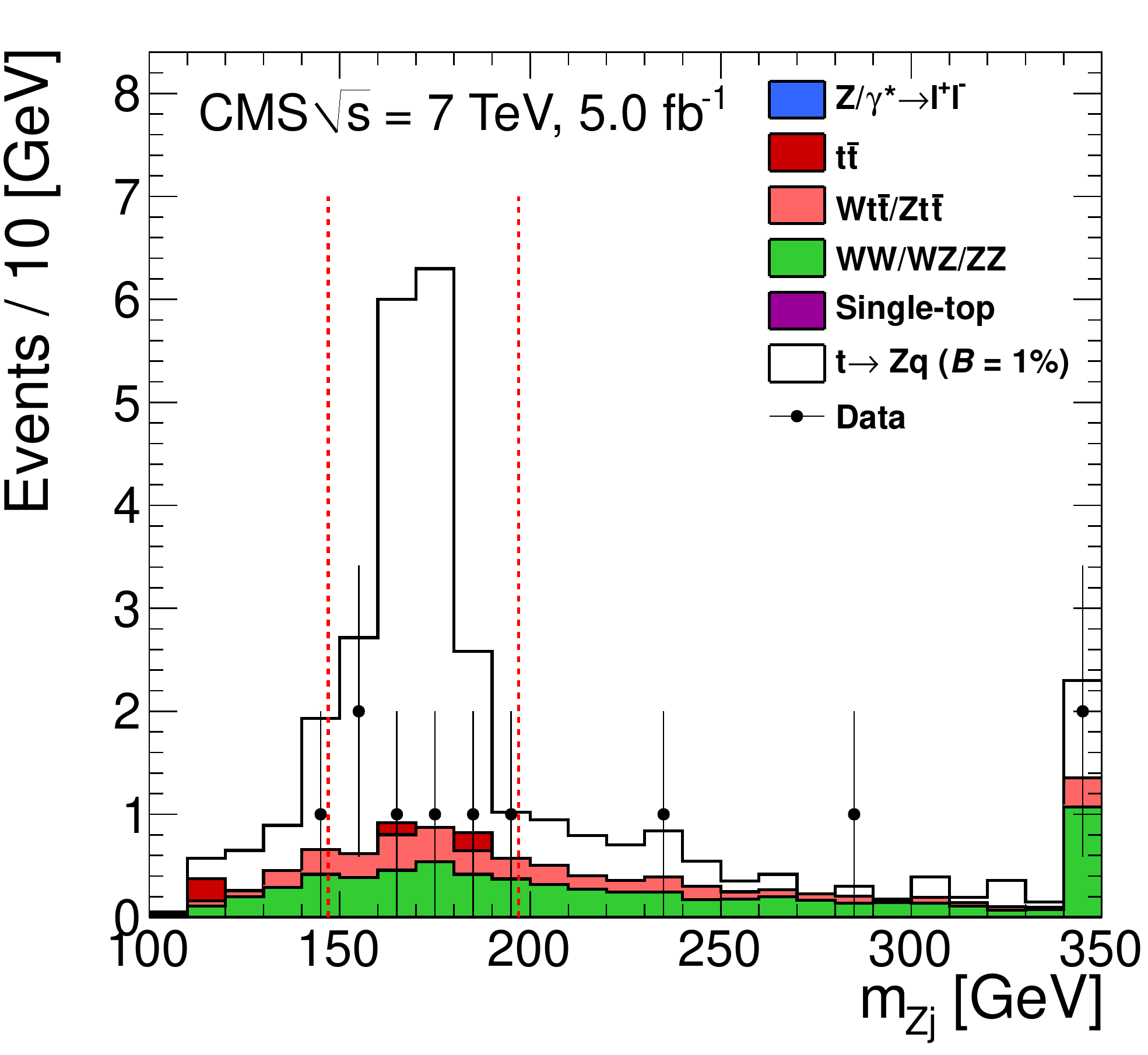}
\includegraphics[width=0.45\textwidth]{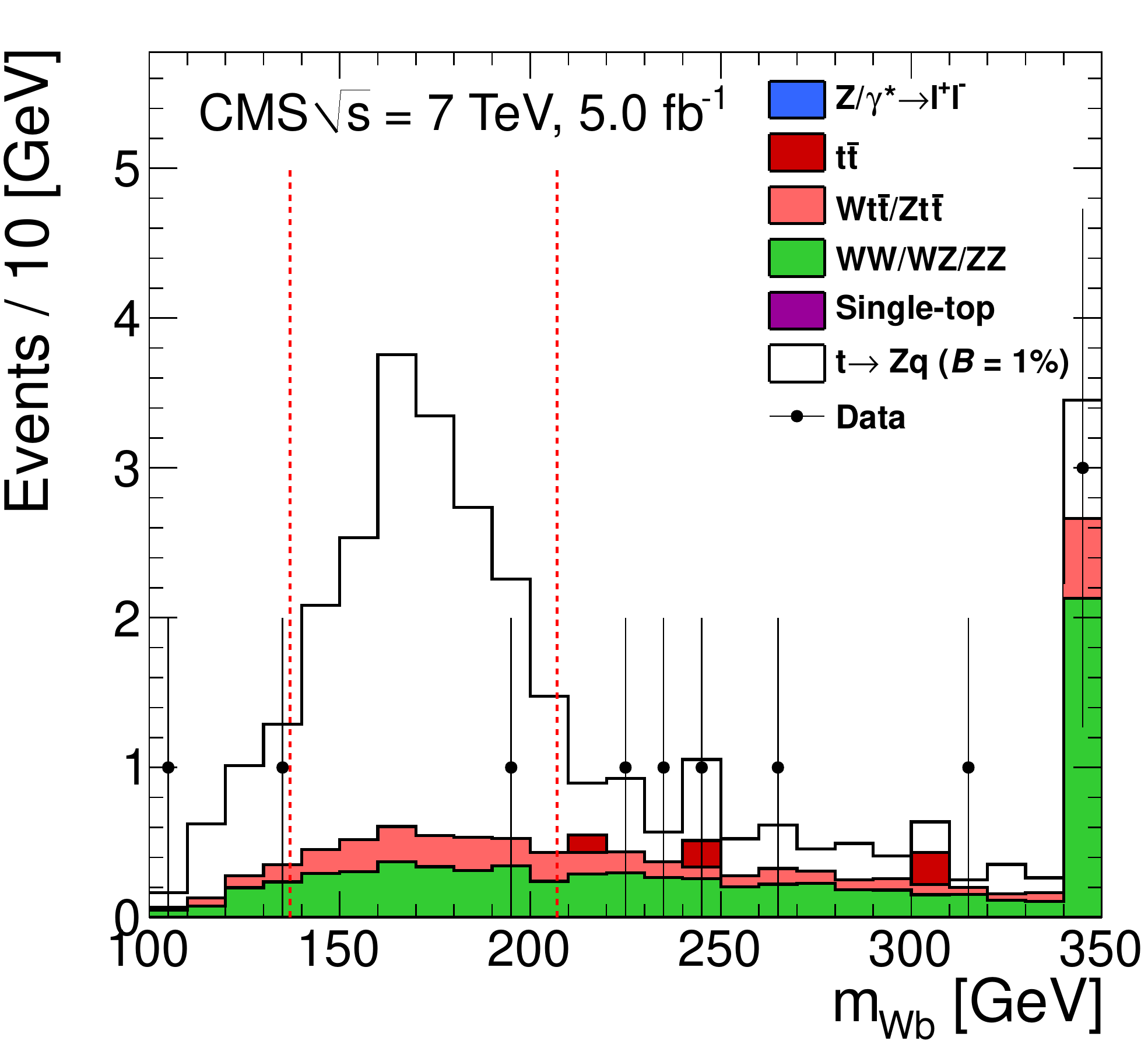}
\caption{\label{fig:mass}
Comparison between data and simulated events of  the $m_{\cPZ \rj}$ and $m_{\PW\cPqb}$  distributions after
the basic event selection described in Section~\ref{sec:preselection}
for an integrated  luminosity of 5.0\fbinv, requiring at least two jets and:
(Top)   the minimum  \st  value, as required in the \st selection;
(Bottom)  exactly one $\cPqb$ jet as required in the $\cPqb$-tag  based selection.
The data are represented by the points with error bars and the open  histogram is the expected signal
assuming $\mathcal{B}(\cPqt\to \cPZ\cPq)$ is equal to 1\%. Stacked solid histograms represent the dominant backgrounds.
The statistical uncertainties of these backgrounds are around a few percent
level and are not drawn.
The last bin contains all the overflow events.
The red dotted lines show the boundaries  of the allowed mass region.
}
\end{center}
\end{figure*}

Figure~\ref{fig:mass}\,(bottom)
shows the comparison between data and simulated events for
$m_{\cPZ \rj}$ and $m_{\PW\cPqb}$    after the basic event
selection and requiring at least two jets, one of which is a  $\cPqb$ jet.

\section{Background Estimation}
\label{sec:back}

Backgrounds are estimated from the yields of simulated events passing the full
selection for $\PW\PW$,  $\PW\cPZ$,  $\cPZ\cPZ$, $\PW\ttbar$,  $\cPZ\ttbar$,
and single-top-quark production, while estimates based on  data are made for
Drell--Yan and $\ttbar$ backgrounds.
The uncertainties in the background estimation given below include, in order,
the statistical and systematic components.

The $\PW\cPZ$ and $\cPZ\cPZ$  production are the dominant diboson  backgrounds.
The production of $\PW$ pairs has a higher cross section, but is unlikely
to contain both an extra high-\pt lepton and a $\cPqb$ jet.
The diboson background estimates  are
$13.6\pm 0.2  \pm 2.6  $ ($ 0.72\pm 0.01 \pm 0.15$)
and
$1.09\pm 0.02\pm 0.21$ ($0.058\pm 0.001 \pm 0.012$)
for the $\PW\cPZ$ and $\cPZ\cPZ$ processes in the \st ($\cPqb$-tag) selection.
These estimates have been rescaled by $1.3 \pm 0.1$ to take into account the
overall normalization difference observed between
data and simulation for the zero jet events, after the event selection given in
Section~\ref{sec:preselection}. The uncertainty of the rescaling factor
estimated from statistical fluctuations in the data contributes to the
systematic uncertainties on the diboson background estimates.
The single-top-quark background is smaller than $0.01$ at the 95\% CL
in both selections.

The  $\cPZ\ttbar$ and $\PW\ttbar$ cross sections are of the same order~\cite{ref:CMS_Vtt}.  The corresponding  background estimates  are
$3.75\pm 0.06 \pm 2.30 $ ($ 0.260\pm 0.004\pm 0.160$)
and
$0.54\pm 0.03\pm 0.36$ ($0.039\pm 0.002 \pm 0.026$)
for the $\cPZ\ttbar$ and $\PW\ttbar$ processes in the \st ($\cPqb$-tag) selection.

The Drell--Yan  background is small due to the minimum 30\GeV requirement of
missing transverse energy. Other backgrounds from QCD multijet events
in which a jet could be misidentified as a lepton are negligible. It is
possible for SM $\ttbar$ to satisfy the $\cPZ$ selection
when both $\PW$ bosons decay leptonically  into the same flavor,
but the third lepton   and  the top quark mass  requirements will reject
these events.

The Drell--Yan and $\ttbar$ background estimates are  based on two data
samples. The first sample is composed of all events satisfying the basic event
selection with two or more jets and loose requirements in \st,
$m_{\cPZ \rj}$, and  $m_{\PW\cPqb}$. The second sample also has  loose requirements
in \st, $m_{\cPZ \rj}$,
and  $m_{\PW\cPqb}$, but in addition it also has  a less stringent isolation criteria for the third lepton.
Therefore,  the second sample is an admixture of the purer three-lepton sample plus events with a misidentified lepton,
originating  from jets or heavy-flavor decays, or genuine three-lepton events that were lost  in the signal sample
due to the more stringent isolation requirement.
The number of events in the two samples  are then  related by the  efficiency of  events
with nominal lepton isolation   and the probability of a jet to be misidentified as a lepton. Using the genuine
and misidentified lepton
efficiencies, which are both determined from data, the yield  of genuine and misidentified three-lepton events is found.
This measurement is turned into an estimate of the  Drell--Yan and $\ttbar$
background after subtracting the contribution from
dibosons  and taking into account the change in acceptances and  efficiencies ({$\eg$} $\cPqb$-tagging)
after the full signal selections are made.
The total contribution of Drell--Yan and  $\ttbar$ events, after the
\st  and $\cPqb$-tag  selections
are estimated to be $1.5\pm 0.5 \pm 0.4$  and $0.06\pm 0.02 \pm 0.01$,
respectively. The statistical and systematic uncertainties
are estimated from the amounts of the events of these two data samples
and the uncertainty of the lepton isolation efficiencies measured with data.
These estimates are compatible with the  expectations based on simulated events.
The total estimated backgrounds are given in Table~\ref{tab:yields}.

\begin{table*}[htb]
\begin{center}
\topcaption{
\label{tab:yields}
Background composition,
observed and expected yields, and limits at the
95\% CL  for all  three-lepton channels combined
for  the \st and $\cPqb$-tag
selections for an  integrated luminosity of 5.0\fbinv.
The uncertainties in the background estimation include the statistical and systematic components
separately (in that order).
}
\begin{tabular}{| l | c | c |}
   \hline
\hline
Selection  & \st & $\cPqb$-tag \\
\hline
\hline
 $\PW\cPZ$ background & $13.59\pm 0.20 \pm 2.58 $ &$ 0.718\pm 0.011\pm 0.150$\\
 $\cPZ\cPZ$ background & $1.09\pm 0.02\pm 0.21$   &$0.058\pm 0.001 \pm 0.012$\\
Drell-Yan and  $\ttbar$ background &$1.52\pm 0.46\pm 0.41$  &   $0.055\pm 0.017 \pm 0.012$\\
 $\cPZ\ttbar$ background &$3.75\pm 0.06\pm 2.30$  &   $0.260\pm 0.004 \pm 0.160$\\
 $\PW\ttbar$ background &$0.54\pm 0.03\pm 0.36$  &   $0.039\pm 0.002 \pm 0.026$\\
\hline
 Total background prediction & 20.49$\pm$ 0.51 $\pm$ 3.51 & 1.13 $\pm$ 0.02  $\pm$ 0.22 \\
 Observed events       & 11     & 0 \\
 \hline
 Expected limit at the 95\% CL  &  ${\cal B}(\cPqt\to\cPZ\cPq)< 0.40\%$      & ${\cal B}(\cPqt\to\cPZ\cPq)< 0.41\%$ \\
 Observed limit at the 95\% CL  &  ${\cal B}(\cPqt\to\cPZ\cPq)< 0.21\%$      & ${\cal B}(\cPqt\to\cPZ\cPq)< 0.30\%$ \\
\hline
\end{tabular}
\end{center}
\end{table*}

\section{Systematic Uncertainties}
\label{sec:sys}
The systematic uncertainties come from the trigger efficiency, choice of  parton   distribution functions, lepton selection, pileup  modeling,
missing transverse energy resolution,
 uncertainty on the $\ttbar$ cross section and diboson rescaling,
$\cPqb$-tagging efficiency for high-$\pt$ $\cPqb$ jets~\cite{ref:btag}, and
jet energy scale~\cite{ref:jetscale}. The prescription given in~\cite{ref:cteq}
is  used to determine the uncertainty
from the choice of parton distribution functions.

In addition,
there is a 2.2\% uncertainty on the luminosity measurement~\cite{ref:lumi2012}.
All these sources combine to give  a 19\%\,(21\%) relative uncertainty on the signal acceptance times efficiency in the
\st ($\cPqb$-tag) selection.

The systematic uncertainties  are summarized in Table~\ref{tab:sys}.
The systematic uncertainty of the background estimation is listed with the total
background prediction given in Table~\ref{tab:yields}.

\begin{table*}[htb]
\begin{center}
\topcaption{
\label{tab:sys}
Summary of the systematic uncertainties for the event selection in percent for  the \st and $\cPqb$-tag
selections. There is an additional 2.2\% uncertainty due to the luminosity measurement.}
\begin{tabular}{| l | c | c|}
   \hline
\hline
Source     & \st selection $[\%]$  & $\cPqb$-tag selection $[\%]$\\
   \hline
   \hline
Trigger efficiency &  4&  4 \\
Parton   distribution functions & 6 & 6\\
Lepton selection  &  7  &  7\\
Pileup events  & 7 & 7 \\
Missing transverse energy resolution  & 8 & 8\\
Cross sections and rescaling & 8 & 8\\
$\cPqb$ tagging &  ---  & 9 \\
Jet energy scale  & 10 & 10\\
\hline
\hline
Total &19 &21 \\
\hline
\hline
\end{tabular}
\end{center}
\end{table*}

\section{Results}
\label{sec:results}
In the \st ($\cPqb$-tag) selection,  we expect $20.5\pm 3.5$  ($1.1 \pm 0.2$)
events from the SM background processes and we observe 11\,(0) events for all
four channels combined.
When all statistical and systematic uncertainties are taken into account,
the probability for the expected number of events, 20.5,
to fluctuate to 11 events, as observed, or fewer is 5\%.
No excess beyond the SM background is observed and a 95\% CL upper limit on
the branching fraction of $\cPqt\to\cPZ\cPq$ is determined
using the modified frequentist approach (CL$_\mathrm{s}$
method~\cite{ref:Junk:1999kv,ref:Read:2002hq}).
A summary of the observed and predicted yields and limits are presented in
Table~\ref{tab:yields}.

The calculation of the upper limit is based on
the information provided by the observed event count
combined with the values and the uncertainties of the luminosity measurement, the background prediction,
and the fraction of all $\ttbar\to \cPZ\cPq +  \PW\cPqb \to  \ell \ell \cPq  + \ell \cPqb$ events expected to be
selected. The signal event yield is
obtained from the efficiency times  acceptance and  branching fraction for simulated events.
As $\mathcal{B}(\cPqt \to \cPZ\cPq)$ is expected to be small,
the possibility of both top quarks decaying via flavor changing neutral currents is not considered.

The best observed  and expected 95\% CL upper limits
on the branching fraction ${\cal B}(\cPqt \to \cPZ\cPq)$ are 0.21\% and 0.40\%, respectively,
obtained in the \st selection from the combined three-lepton analyses.
The one-sigma boundaries of the expected limit are
0.30--0.59\%.  The corresponding observed and expected upper
limits, and one-sigma boundaries for the $\cPqb$-tag  selection
are  0.30\%\,, 0.41\% and 0.30--0.53\%, respectively.
The  expected limit for  the \st and $\cPqb$-tag selections show that they have comparable
sensitivity.  The one with slightly better expected limit is taken as the final result.

\section{Summary}
\label{sec:conclusions}
A search for flavor changing neutral currents in top quark decays in
$\ttbar$ events
produced in proton-proton collisions at $\sqrt{s} = 7$\TeV is presented.
A sample of three-lepton events is selected from data recorded by CMS during
2011 corresponding to an integrated luminosity of 5.0\fbinv.
These events are compatible with  a
$\Pp\Pp \to \ttbar \to$  $\cPZ\cPq + \PW\cPqb \to$ $\ell\ell \cPq +  \ell \nu\cPqb\,( \ell = \Pe,\,\mu)$ topology.
Since three-lepton events originating from the SM processes are rare
the background contributions are small. No excess of events over the SM background is observed and a
 ${\cal B}({\cPqt\to\cPZ\cPq})$
 branching fraction larger than 0.21\% is excluded at the 95\% confidence level.\\\\

\section*{Acknowledgments}
We congratulate our colleagues in the CERN accelerator departments for the excellent
performance of the LHC machine. We thank the technical and administrative staff at CERN
and other CMS institutes, and acknowledge support from: FMSR (Austria); FNRS and FWO
(Belgium); 	CNPq, CAPES, FAPERJ, and FAPESP (Brazil); MES (Bulgaria); CERN; CAS,
MoST, and NSFC (China); COLCIENCIAS (Colombia); MSES (Croatia); RPF (Cyprus); MoER,
SF0690030s09 and RDF (Estonia); Academy of Finland, MEC, and HIP (Finland); CEA and
CNRS/IN2P3 (France); BMBF, DFG, and HGF (Germany); GSRT (Greece); OTKA and NKTH
(Hungary); DAE and DST (India); IPM (Iran); SFI (Ireland); INFN (Italy); NRF and WCU
(Korea); LAS (Lithuania); CINVESTAV, CONACYT, SEP, and UASLP-FAI (Mexico); MSI
(New Zealand); PAEC (Pakistan); MSHE and NSC (Poland); FCT (Portugal); JINR (Armenia,
Belarus, Georgia, Ukraine, Uzbekistan); MON, RosAtom, RAS and RFBR (Russia); MSTD
(Serbia); MICINN and CPAN (Spain); Swiss Funding Agencies (Switzerland); NSC (Taipei);
TUBITAK and TAEK (Turkey); STFC (United Kingdom); DOE and NSF (USA).

\bibliography{auto_generated}   

\providecommand{\href}[2]{#2}\begingroup\raggedright\begin{thebibliography}{10}%
\makeatletter
\providecommand{\hrefCMSnoop }[0]{\@secondoftwo}%
\makeatother
\providecommand{\doi}{\texttt{doi:}\begingroup \urlstyle{tt}\Url}

\bibitem{ref:GIM}
\hrefCMSnoop {} {S.~L. Glashow, J.~Iliopoulos, and L.~Maiani, ``Weak
  Interactions with Lepton-Hadron Symmetry'',} \textit{ Phys. Rev. D} \textbf{
  2} (1970) 1285,
  \href{http://dx.doi.org/10.1103/PhysRevD.2.1285}{\doi{10.1103/PhysRevD.2.1285}}.

\bibitem{ref:Glover:2004cy}
\href {http://th-www.if.uj.edu.pl/acta/vol35/abs/v35p2671.htm} {J.~A.
  Aguilar-Saavedra {et~al.}, ``{Top quark physics}'',} \textit{ Acta Phys.
  Polon. B} \textbf{ 35} (2004) 2671,
\href{http://www.arXiv.org/abs/hep-ph/0410110}{\texttt{ arXiv:hep-ph/0410110}}.

\bibitem{AguilarSaavedra:2000db}
\hrefCMSnoop {} {J.~A. Aguilar-Saavedra, ``{Top flavor changing neutral
  coupling signals at a linear collider}'',} \textit{ Phys. Lett. B} \textbf{
  502} (2001) 115,
  \href{http://dx.doi.org/10.1016/S0370-2693(01)00162-9}{\doi{10.1016/S0370-2693(01)00162-9}},
\href{http://www.arXiv.org/abs/hep-ph/0012305}{\texttt{ arXiv:hep-ph/0012305}}.

\bibitem{Lu:2003yr}
G.~Lu\hrefCMSnoop {} { {et~al.}, ``{Rare top quark decays $t\to cV$ in the
  top-color-assisted technicolor model}'',} \textit{ Phys. Rev. D} \textbf{ 68}
  (2003) 015002,
  \href{http://dx.doi.org/10.1103/PhysRevD.68.015002}{\doi{10.1103/PhysRevD.68.015002}},
\href{http://www.arXiv.org/abs/hep-ph/0303122}{\texttt{ arXiv:hep-ph/0303122}}.

\bibitem{ref:2008aaa}
\hrefCMSnoop {} {{ CDF} Collaboration, ``{Search for the flavor changing
  neutral current decay t $\to \rm Zq$ in ${\rm p \bar{p}}$ collisions at
  $\sqrt{s}=1.96$ TeV}'',} \textit{ Phys. Rev. Lett.} \textbf{ 101} (2008)
  192002,
  \href{http://dx.doi.org/10.1103/PhysRevLett.101.192002}{\doi{10.1103/PhysRevLett.101.192002}},
\href{http://www.arXiv.org/abs/0805.2109}{\texttt{ arXiv:0805.2109}}.

\bibitem{ref:Abazov:2011qf}
\hrefCMSnoop {} {{ D0} Collaboration, ``{Search for flavor changing neutral
  currents in decays of top quarks}'',} \textit{ Phys. Lett. B} \textbf{ 701}
  (2011) 313,
  \href{http://dx.doi.org/10.1016/j.physletb.2011.06.014}{\doi{10.1016/j.physletb.2011.06.014}},
\href{http://www.arXiv.org/abs/1103.4574}{\texttt{ arXiv:1103.4574}}.

\bibitem{ref:ATLAS-2012}
\hrefCMSnoop {} {{ ATLAS} Collaboration, ``{A search for flavour changing
  neutral currents in top-quark decays in pp collision data collected with the
  ATLAS detector at sqrt(s) = 7 TeV}'',} \textit{ JHEP} \textbf{ 1209} (2012)
  139,
  \href{http://dx.doi.org/10.1007/JHEP09(2012)139}{\doi{10.1007/JHEP09(2012)139}},
\href{http://www.arXiv.org/abs/1206.0257}{\texttt{ arXiv:1206.0257}}.

\bibitem{ref:cms}
\hrefCMSnoop {} {{ CMS} Collaboration, ``The {CMS} experiment at the {CERN
  LHC}'',} \textit{ JINST} \textbf{ 03} (2008) S08004,
\href{http://dx.doi.org/10.1088/1748-0221/3/08/S08004}{\doi{10.1088/1748-0221/3/08/S08004}}.

\bibitem{Khachatryan:2010xn}
\hrefCMSnoop {} {{ CMS} Collaboration, ``{Measurements of inclusive W and Z
  cross sections in pp collisions at $\sqrt{s}$ = 7 TeV}'',} \textit{ JHEP}
  \textbf{ 01} (2011) 080,
  \href{http://dx.doi.org/10.1007/JHEP01(2011)080}{\doi{10.1007/JHEP01(2011)080}},
\href{http://www.arXiv.org/abs/1012.2466}{\texttt{ arXiv:1012.2466}}.

\bibitem{Chatrchyan:2012xi}
\hrefCMSnoop {} {{ CMS} Collaboration, ``{Performance of CMS muon
  reconstruction in pp collision events at sqrt(s) = 7 TeV}'',} \textit{ JINST}
  \textbf{ 07} (2012) P10002,
  \href{http://dx.doi.org/10.1088/1748-0221/7/10/P10002}{\doi{10.1088/1748-0221/7/10/P10002}},
\href{http://www.arXiv.org/abs/1206.4071}{\texttt{ arXiv:1206.4071}}.

\bibitem{ref:pf}
\href {http://cdsweb.cern.ch/record/1194487} {{ CMS} Collaboration,
  ``Particle--Flow Event Reconstruction in {CMS} and Performance for Jets,
  Taus, and {\MET}'',} CMS Physics Analysis Summary CMS-PAS-PFT-09-001, (2009).

\bibitem{ref:kt}
\hrefCMSnoop {} {M.~Cacciari and G.~P. Salam, ``{Dispelling the $N^{3}$ myth
  for the $k_t$ jet-finder}'',} \textit{ Phys. Lett. B} \textbf{ 641} (2006)
  57,
  \href{http://dx.doi.org/10.1016/j.physletb.2006.08.037}{\doi{10.1016/j.physletb.2006.08.037}},
  \href{http://www.arXiv.org/abs/hep-ph/0512210}{\texttt{
  arXiv:hep-ph/0512210}}.

\bibitem{ref:jetscale}
\hrefCMSnoop {} {{ CMS} Collaboration, ``{Determination of jet energy
  calibration and transverse momentum resolution in CMS}'',} \textit{ JINST}
  \textbf{ 06} (2011) P11002,
  \href{http://dx.doi.org/10.1088/1748-0221/6/11/P11002}{\doi{10.1088/1748-0221/6/11/P11002}},
\href{http://www.arXiv.org/abs/1107.4277}{\texttt{ arXiv:1107.4277}}.

\bibitem{ref:met}
\hrefCMSnoop {} {{ CMS} Collaboration, ``{Missing transverse energy performance
  of the CMS detector}'',} \textit{ JINST} \textbf{ 06} (2011) P09001,
  \href{http://dx.doi.org/10.1088/1748-0221/6/09/P09001}{\doi{10.1088/1748-0221/6/09/P09001}},
\href{http://www.arXiv.org/abs/1106.5048}{\texttt{ arXiv:1106.5048}}.

\bibitem{ref:noise}
\hrefCMSnoop {} {{ CMS} Collaboration, ``Identification and filtering of
  uncharacteristic noise in the {CMS} hadron calorimeter'',} \textit{ JINST}
  \textbf{ 05} (2010) T03014,
  \href{http://dx.doi.org/10.1088/1748-0221/5/03/T03014}{\doi{10.1088/1748-0221/5/03/T03014}},
  \href{http://www.arXiv.org/abs/0911.4881}{\texttt{ arXiv:0911.4881}}.

\bibitem{ref:madg}
J.~Alwall\hrefCMSnoop {} { {et~al.}, ``{MadGraph 5}: going beyond'',} \textit{
  JHEP} \textbf{ 06} (2011) 128,
  \href{http://dx.doi.org/10.1007/JHEP06(2011)128}{\doi{10.1007/JHEP06(2011)128}},
\href{http://www.arXiv.org/abs/1106.0522}{\texttt{ arXiv:1106.0522}}.

\bibitem{ref:pythia}
\hrefCMSnoop {} {T.~Sj{\"o}strand, S.~Mrenna, and P.~Skankds, ``{PYTHIA} 6.4
  physics and manual'',} \textit{ JHEP} \textbf{ 05} (2006) 026,
  \href{http://dx.doi.org/10.1088/1126-6708/2006/05/026}{\doi{10.1088/1126-6708/2006/05/026}},
\href{http://www.arXiv.org/abs/hep-ph/0603175}{\texttt{ arXiv:hep-ph/0603175}}.

\bibitem{ref:Nason:2004rx}
\hrefCMSnoop {} {P.~Nason, ``{A new method for combining NLO QCD with shower
  Monte Carlo algorithms}'',} \textit{ JHEP} \textbf{ 11} (2004) 040,
  \href{http://dx.doi.org/10.1088/1126-6708/2004/11/040}{\doi{10.1088/1126-6708/2004/11/040}},
\href{http://www.arXiv.org/abs/hep-ph/0409146}{\texttt{ arXiv:hep-ph/0409146}}.

\bibitem{ref:Frixione:2007vw}
\hrefCMSnoop {} {S.~Frixione {et~al.}, ``{Matching NLO QCD computations with
  parton shower simulations: the POWHEG method}'',} \textit{ JHEP} \textbf{ 11}
  (2007) 070,
  \href{http://dx.doi.org/10.1088/1126-6708/2007/11/070}{\doi{10.1088/1126-6708/2007/11/070}},
\href{http://www.arXiv.org/abs/0709.2092}{\texttt{ arXiv:0709.2092}}.

\bibitem{ref:Alioli:2010xd}
S.~Alioli\hrefCMSnoop {} { {et~al.}, ``{A general framework for implementing
  NLO calculations in shower Monte Carlo programs: the POWHEG BOX}'',} \textit{
  JHEP} \textbf{ 06} (2010) 043,
  \href{http://dx.doi.org/10.1007/JHEP06(2010)043}{\doi{10.1007/JHEP06(2010)043}},
\href{http://www.arXiv.org/abs/1002.2581}{\texttt{ arXiv:1002.2581}}.

\bibitem{ref:CTEQ6L}
P.~M. Nadolsky\hrefCMSnoop {} { {et~al.}, ``Implications of CTEQ global
  analysis for collider observables'',} \textit{ Phys. Rev. D} \textbf{ 78}
  (2008) 013004,
  \href{http://dx.doi.org/10.1103/PhysRevD.78.013004}{\doi{10.1103/PhysRevD.78.013004}}.

\bibitem{ref:geant}
\hrefCMSnoop {} {{ GEANT4} Collaboration, ``{GEANT4}---a simulation toolkit'',}
  \textit{ Nucl. Instrum. Meth. A} \textbf{ 506} (2003) 250,
\href{http://dx.doi.org/10.1016/S0168-9002(03)01368-8}{\doi{10.1016/S0168-9002(03)01368-8}}.

\bibitem{ref:btag}
\href {http://cdsweb.cern.ch/record/1366061} {{ CMS} Collaboration,
  ``Performance of b-jet identification in {CMS}'',} CMS Physics Analysis
  Summary CMS-PAS-BTV-11-001, (2011).

\bibitem{ref:CMS_Vtt}
\href {http://cdsweb.cern.ch/record/1478935} {{ CMS} Collaboration,
  ``Measurement of the Associated Production of Vector Bosons with Top-Antitop
  Pairs at 7 TeV'',} CMS Physics Analysis Summary CMS-PAS-TOP-12-014, (2012).

\bibitem{ref:cteq}
\hrefCMSnoop {} {M.~R. Whalley, D.~Bourilkov, and R.~C. Group, ``{The Les
  Houches accord PDFs (LHAPDF) and LHAGLUE}'',} (2005).
\href{http://www.arXiv.org/abs/hep-ph/0508110}{\texttt{ arXiv:hep-ph/0508110}}.

\bibitem{ref:lumi2012}
\href {http://cdsweb.cern.ch/record/1434360} {{ CMS} Collaboration, ``Absolute
  Calibration of the Luminosity Measurement at {CMS}: {W}inter 2012 Update'',}
  CMS Physics Analysis Summary CMS-PAS-SMP-12-008, (2012).

\bibitem{ref:Junk:1999kv}
\hrefCMSnoop {} {T.~Junk, ``{Confidence level computation for combining
  searches with small statistics}'',} \textit{ Nucl. Instrum. Meth. A} \textbf{
  434} (1999) 435,
  \href{http://dx.doi.org/10.1016/S0168-9002(99)00498-2}{\doi{10.1016/S0168-9002(99)00498-2}},
\href{http://www.arXiv.org/abs/hep-ex/9902006}{\texttt{ arXiv:hep-ex/9902006}}.

\bibitem{ref:Read:2002hq}
\hrefCMSnoop {} {A.~L. Read, ``{Presentation of search results: the $CL_s$
  technique}'',} \textit{ J. Phys. G} \textbf{ 28} (2002) 2693,
\href{http://dx.doi.org/10.1088/0954-3899/28/10/313}{\doi{10.1088/0954-3899/28/10/313}}.

\end{thebibliography}\endgroup

\cleardoublepage \appendix\section{The CMS Collaboration \label{app:collab}}\begin{sloppypar}\hyphenpenalty=5000\widowpenalty=500\clubpenalty=5000\textbf{Yerevan Physics Institute,  Yerevan,  Armenia}\\*[0pt]
S.~Chatrchyan, V.~Khachatryan, A.M.~Sirunyan, A.~Tumasyan
\vskip\cmsinstskip
\textbf{Institut f\"{u}r Hochenergiephysik der OeAW,  Wien,  Austria}\\*[0pt]
W.~Adam, E.~Aguilo, T.~Bergauer, M.~Dragicevic, J.~Er\"{o}, C.~Fabjan\cmsAuthorMark{1}, M.~Friedl, R.~Fr\"{u}hwirth\cmsAuthorMark{1}, V.M.~Ghete, J.~Hammer, N.~H\"{o}rmann, J.~Hrubec, M.~Jeitler\cmsAuthorMark{1}, W.~Kiesenhofer, V.~Kn\"{u}nz, M.~Krammer\cmsAuthorMark{1}, D.~Liko, I.~Mikulec, M.~Pernicka$^{\textrm{\dag}}$, B.~Rahbaran, C.~Rohringer, H.~Rohringer, R.~Sch\"{o}fbeck, J.~Strauss, A.~Taurok, P.~Wagner, W.~Waltenberger, G.~Walzel, E.~Widl, C.-E.~Wulz\cmsAuthorMark{1}
% ...
\vskip\cmsinstskip
\dag:~Deceased\\
1:~~Also at Vienna University of Technology, Vienna, Austria\\
% ...

\end{sloppypar}
\end{document}